\documentclass[floats,twocolumn,prb]{revtex4}
\usepackage{graphicx}% Include figure files
\usepackage{dcolumn}% Align table columns on decimal point
\usepackage{bm}% bold math
\usepackage{mathrsfs}

\begin{document}
%\citestyle{nature}
\preprint{APS/123-QED}

\title{Evolution between two orbital-selective Mott phases driven by interorbital hopping}
% Force line breaks with \\

\author{Yu Ni$^1$}
\author{Jian Sun$^2$}
\author{Ya-Min Quan$^3$}
\author{Yun Song$^1$}\emph{}
\thanks{yunsong@bnu.edu.cn}

\affiliation{%
$^1$Department of Physics, Beijing Normal University, Beijing 100875, China}%

\affiliation{%
$^2$Department of Physics, Science and Technology University, Shanghai 201210, China}%

\affiliation{%
$^3$Key Laboratory of Materials Physics, Institute of Solid State Physics,
 Chinese Academy of Sciences, P. O. Box 1129, Hefei 230031, China}%

\date{\today}

\begin{abstract}
  The effect of interorbital hopping on the orbital selective Mottness in a two-band
  correlation system is investigated by using the dynamical mean-field theory with
  the Lanczos method as impurity solver.
  We construct the phase diagram of the two-orbital Hubbard model with interorbital hopping ($t_{12})$,
  where the orbital selective Mott phases (OSMP) show different evolution trends.
  We find that the negative interorbital hopping ($t_{12}<0$) can enhance the OSMP regime
  upon tuning the effective bandwidth ratio.
  On the contrary, for the cases with positive interorbital hopping ($t_{12}>0$), the OSMP region
  becomes narrow with the increase of orbital hybridization until it disappears.
  It is also shown that a new OSMP emerges for a large enough positive interorbital hopping,
  owing to the role exchange of wide and narrow effective orbitals caused by the large $t_{12}$.
  Our results are also applicable to the hole-overdoped Ba$_2$CuO$_{4-\delta}$ superconductor, which is
  an orbital-selective Mott compound at half-filling.
\end{abstract}

\pacs{ 71.27.+a, 71.30.+h, 74.72.-h}
%71.27.+a Strongly correlated electron systems; heavy fermions
%7 1.30.+h Metal-insulator transitions and other electronic transitions
%74.72.-h Cuprate superconductors (high-Tc and insulating parent compounds

\maketitle

%\tableofcontents

%\section{INTRODUCTION}
%\label{sec:INTR}

\section{INTRODUCTION}

The orbital selective Mottness is helpful for exploring the nature of strong correlation systems
due to its discovery in some transition metal compounds, such as
Ca$_{2-x}$Sr$_x$RuO$_4$\cite{Anisimov-2002}, transition metal dichalcognide\cite{Zhicheng Zhong-2017},
and Fe-based superconductors\cite{Jianwei Huang-2021}.
When the carries on a subset of orbitals get localized while the others remain itinerate,
the orbital-selective Mott transition (OSMT) happens.
The simplest theoretical realization
of the OSMT occurs in the two-orbital Hubbard model\cite{oles-1983,Koga-2002,Koga-2005,Kubo-2007}.
The dynamical mean-field theory \cite{Georges-1996,Held-2001,Kotliar-2006} (DMFT) is a powerful framework
to study the correlation-driven phase transitions in the one-band Hubbard model,
and its extension to the two-orbital Hubbard model is also effective\cite{Koga-2002,Liebsch-2005,Peters-2011}.

Several theoretical approaches have been used to build the impurity solver of the DMFT procedure,
such as quantum Monte Carlo simulations
(QMC)\cite{Hirsch-1983,Hirsch-1985,Chang-2008,Tocchio-2016},
renormalization-group theory\cite{Hanke-1982,Shankar-1994,Hille-2020},
and slave-variable representations\cite{Kotliar-1986,Lechermann-2007,Rong Yu-2010}, etc.
For the multi-orbital extensions combined with the DMFT algorithm, each solver has its limitations.
The QMC method faces the
sign problem in the doped fermion system, renormalization-group
theory can solve the one-band model well but it is hard to be
expanded to multi-band system, and slave-variable representations can not
treat the interaction effect accurately.  For the multi-orbital correlation
system, the DMFT with the Lanczos method as impurity solver can
accurately treat the multi-orbital correlations including the intraorbital
interaction $U$, interorbital correlation $U^{\prime}$,
and Hund's rule coupling $J_H$. But the off-diagonal
Green's function induced by the interorbital hopping will add
complexity of self-consistency and need a lot of computing resources.

The interobital hoppings induce strong orbital hybridization, which is
crucial in many multi-band correlated transition-metal compounds
\cite{Imada-1998,Wen-2006,Rohringer-2018}.
However, it is a tremendous challenge to solve the extend multi-band Hubbard model
which also has the off-diagonal Hamiltonian induced by the interorbital hopping.
We introduce the canonical transformation to diagonalize the tight-binding part
of the extended two-orbital Hubbard Hamiltonian\cite{Song-2005,Song-2009},
and the effective orbitals obtained can reflect the effect of orbital hybridization
in the multiorbital correlation system.
To comprehensively study the cooperate effects of multiorbital interactions,
we also develop the Lanczos method as the DMFT impurity solver.
Comparing with the previous modified DMFT procedure\cite{Song-2005,Song-2009},
our present work can treat the Coulomb interactions and Hond's rule coupling strictly, especially the critical
points of the phase transitions can be determined accurately.
We use the DMFT with Lanczos solver to study the electron correlation effect of two-orbital
Hubbard model with sign convertable interorbital hopping, and we find two OSMP
regions in the phase diagram. We also apply our results to analyze the recently discovered
two-orbital superconductor Ba$_2$CuO$_{4-\delta}$\cite{JinCQ-PNAS-2019}.
We find the orbital-select Mott transition in the half-filled
Ba$_2$CuO$_{4-\delta}$.

This paper is organized as follows. In Sec. II we introduce
the canonical transformation used for the two-orbital Hubbard
model including interorbital hopping. In Sec. III we explain
the numerical method adopted to solve the transformed effective
model: the DMFT approach with Lanczos solver. In Sec. IV
we present the results of the orbital-selective Mott transitions in the two-orbital Hubbard model, and
discuss the  cooperate effect of electron correlation and interorbital
hopping in the multi-band correlation system. In Sec. V we apply the method to the
two-orbital superconductor Ba$_2$CuO$_{3.5}$ and introduce
our finding. The principal conclusions of this paper are
summarized in Sec. VI.

%^^^^^^^^^^^^^^^^^^^^^^^^^^^^^^^^^^^^^^^^^%
\section{Canonical transformation}

The two-orbital Hamiltonian consists of two parts:
tight-binding Hamiltonian $H_t$ and
interaction Hamiltonian $H_I$ ,
where the tight-binding Hamiltonian $H_t$ reads
\begin{eqnarray}
    H_t&=&-\sum_{\langle ij \rangle}\sum_{l \sigma}
    t_ld^{\dag}_{il\sigma}d_{jl\sigma}
    -\sum_{\langle ij \rangle}\sum_{l\neq l', \sigma}
    t_{ll'}d^{\dag}_{il\sigma}d_{jl'\sigma}
    \nonumber\\
    &&-\mu\sum_{il\sigma} d^{\dag}_{il\sigma}d_{il\sigma},
\label{Eq:tightbonding}
\end{eqnarray}
and interaction Hamiltonian $H_I$ is given by\cite{oles-1983,Koga-2005}
\begin{eqnarray}
    H_I&=&\frac{U}{2}\sum_{il\sigma}n_{il\sigma}n_{il\bar{\sigma}}+\sum_{i,l<l',\sigma\sigma'}
    (U'-\delta_{\sigma\sigma'}J_H)n_{il\sigma}n_{il'\sigma'}
    \nonumber\\
    &&+\frac{J_H}{2}\sum_{i,l\neq l',\sigma} d^{\dag}_{il\sigma}d^{\dag}_{il\bar{\sigma}}d_{il'\bar{\sigma}}d_{il'\sigma}
    \nonumber\\
    &&+\frac{J_H}{2}\sum_{i,l\neq l',\sigma\sigma'} d^{\dag}_{il\sigma}d^{\dag}_{il'\sigma'}d_{il\sigma'}d_{il'\sigma},
\label{Eq:TOHub}
\end{eqnarray}
where $d^{\dag}_{il\sigma}$ ($d_{il\sigma}$) is an electron
creation (annihilation) operator for orbital $l$
at site $i$ with spin $\sigma$, and $\langle ij \rangle$
represent nearest neighbor (NN) sites. $t_l$ denotes the NN intraorbital hopping
and $t_{ll'}$ denotes the NN interorbital hopping.
 $U$ ($U'$) corresponds to the intraorbital
(interorbital) interaction,
and $J_H$ is the Hund's rule coupling.
For the systems with spin rotation symmetry,
we have $U=U'+2J_H$.

We introduce two effective decoupled orbitals $\alpha$
and $\beta$ by a canonical transformation, that
decoupling the interorbital hopping\cite{Song-2005,Song-2009}
\begin{eqnarray}
  d_{i1\sigma } &= u\alpha _{i\sigma } + v\beta _{i\sigma }, \nonumber\\
  d_{i2\sigma } &= -v\alpha _{i\sigma } + u\beta _{i\sigma },
\label{Eq:CT_uv}
\end{eqnarray}
with
\begin{eqnarray}
   {{u} = \frac{1}{\sqrt 2}\left( {1 + \sqrt {\frac{{{{\left( {{t_{11}} - {t_{22}}} \right)}^2}}}{{{(t_{12})^2} + {{\left( {{t_{11}} - {t_{22}}} \right)}^2}}}} } \right)^{1/2}},\nonumber\\
    {{v} = \frac{1}{\sqrt 2}\left( {1 - \sqrt {\frac{{{{\left( {{t_{11}} - {t_{22}}} \right)}^2}}}{{{(t_{12})^2} + {{\left( {{t_{11}} - {t_{22}}} \right)}^2}}}} } \right)^{1/2}},\
\label{Eq:uv}
\end{eqnarray}
where $\alpha_{i\sigma}$ and $\beta_{i\sigma}$ are
fermion annihilation operators for the two newly
introduced $\alpha$ and $\beta$ orbitals.
The values of parameters $u$ and $v$ determined
by Eq.~(\ref{Eq:uv}) will make the interorbital hopping
between the $\alpha$ and $\beta$ orbitals vanish.
Through the canonical transformation, the original
two-orbital Hamiltonians shown in Eq.~(\ref{Eq:tightbonding}) and Eq.~(\ref{Eq:TOHub})
are converted into an effective two-orbital Hamiltonian
$H^{eff}$, which consists of the tight-binding part
$H_t^{eff}$ and the interaction part $H_I^{eff}$ for
the two effective orbitals as:
\begin{eqnarray}
  H_t^{eff} &=&  - \sum\limits_{\left\langle {i,j} \right\rangle \sigma }
     {\left( {{t_\alpha }\alpha _{i\sigma }^+ {\alpha _{j\sigma }}
     + {t_\beta }\beta _{i\sigma }^ + {\beta _{j\sigma }}} \right)}
     \nonumber\\
   &&-\mu \sum\limits_{i\sigma}{\left({\alpha _{i\sigma }^
     + {\alpha _{i\sigma }} + \beta _{i\sigma }^ + {\beta _{i\sigma }}} \right)},
\label{Eq:EMWIH}
\end{eqnarray}
and
\begin{eqnarray}
H_I^{eff}&=&\frac{U}{2} \sum_{i \sigma}\left(n_{i \alpha \sigma} n_{i \alpha \bar{\sigma}}+n_{i \beta \sigma} n_{i \beta \bar{\sigma}}\right)\nonumber\\
&&+\sum_{i \sigma \sigma^{\prime}}\left(U^{\prime}-\delta_{\sigma \sigma^{\prime}} J_H\right) n_{i \alpha \sigma} n_{i \beta \sigma^{\prime}}\nonumber\\
 &&+\frac{J_{H}}{2} \sum_{i, \sigma}\left(\alpha_{i \sigma}^{\dagger} \alpha_{i \bar{\sigma}}^{\dagger} \beta_{i \bar{\sigma}} \beta_{i \sigma}+\beta_{i \sigma}^{\dagger} \beta_{i \bar{\sigma}}^{\dagger} \alpha_{i \bar{\sigma}} \alpha_{i \sigma}\right) \nonumber\\
 &&+\frac{J_{H}}{2} \sum_{i, \sigma \sigma^{\prime}} \left(\alpha_{i \sigma}^{\dagger} \beta_{i \sigma^{\prime}}^{\dagger} \alpha_{i \sigma^{\prime}}\beta_{i \sigma}+\beta_{i \sigma}^{\dagger} \alpha_{i \sigma^{\prime}}^{\dagger} \beta_{i \sigma^{\prime}} \alpha_{i \sigma}\right) .\nonumber\\
\label{Eq:HIeff}
\end{eqnarray}
The hopping parameters in the effective model are expressed as
\begin{eqnarray}
t_{\alpha}=t_1 u^{2}+t_2 v^{2}-t_{12}uv\nonumber\\
t_{\beta}=t_1 v^{2}+t_2 u^{2}+t_{12}uv,
\label{Eq:CT}
\end{eqnarray}
according to Eq.~(\ref{Eq:CT_uv}).
The effective interaction $H_I^{eff}$ in Eq.~(\ref{Eq:HIeff})
has a formulation similar to the original interaction terms
when the spin rotation symmetry is kept with $U=U'+2J_H$.

\section{Dynamical mean-field theory}

The canonical transformation decouples the hybridization
of the two orbitals, so that the effective model is comparably easier to
be solved by DMFT.
In the framework of DMFT\cite{Georges-1996}, we map
the lattice Hamiltonian on to an impurity model with
fewer degrees of freedom,
\begin{eqnarray}
    H_{imp}&=&\sum_{m\sigma}\{\epsilon_{m\sigma}^{\alpha}
               c^{\dag}_{\alpha m\sigma} c_{\alpha m\sigma}
             + \epsilon_{m\sigma}^{\beta} c^{\dag}_{\beta m\sigma}c_{\beta m\sigma}\}
             \nonumber\\
             &&+\sum_{m\sigma}V_{m\sigma}^{\alpha}(c^{\dag}_{\alpha  m\sigma}\alpha_{\sigma}
               +\alpha^{\dag}_{\sigma}c_{\alpha m\sigma})
             \nonumber\\
             &&+\sum_{m\sigma}V_{m\sigma}^{\beta}(c^{\dag}_{\beta  m\sigma}\beta_{\sigma}
               +\beta^{\dag}_{\sigma}c_{\beta m\sigma})
             \nonumber\\
        &&-\mu\sum_{\sigma}\alpha^{\dag}_{\sigma}\alpha_{\sigma}
            -\mu\sum_{\sigma}\beta^{\dag}_{\sigma}\beta_{\sigma}
             \nonumber\\
        &&+H^{eff}_{I}(\alpha, \beta),
\label{Eq:IMP}
\end{eqnarray}
where $c^{\dag}_{\gamma m\sigma}$ ($c_{\gamma m\sigma}$)
denotes the creation (annihilation) operator for the 'environmental bath'
lattice of orbital $\gamma$ ($\gamma=\alpha, \beta$),
$\epsilon_{m\sigma}^{\gamma}$ denotes the energy of
the $m$-th 'environmental bath' of orbital $\gamma$,
and $V_{m\sigma}^{\gamma}$ represents the coupling
between the orbital $\gamma$ of the 'impurity site' and
its 'environmental bath'. We take the bath size $n_b=3$ in our work. It has been proved that the critical points of OSMT calculated by DMFT with Lanczos solver in two-orbital Hubbard model\cite{NiuYK-2019} are almost the same when $n_b\geq3$.

By using the canonical transformation, the two orbitals
are nonhybridized. Thus, the Green's function and
self-energy are all diagonal with respect to the orbital,
so that we can calculate the Green's function and the
parameters $V_{m\sigma}^{\gamma}$ and $\epsilon_{m\sigma}^{\gamma}$
independently.
The Weiss function of the impurity model can be obtained
through the parameters of the impurity Hamiltonian by
\begin{eqnarray}
{\cal G}_{0\gamma \sigma}^{-1}\left(i\omega_n\right) = i\omega_n +\mu
-\varepsilon _{\gamma}-\sum_{m}\frac{\left(V_{m\sigma}^{\gamma}\right)^2}
{i\omega_n-\epsilon_{m\sigma}^{\gamma}}.
\end{eqnarray}
Employing the {\it Lanczos} solver, we can obtain
the Green's function $G_{i m p}^{(\gamma)}$
\cite{Dagotto-1994,Caffarel-1994,Capone-2007}, which is
expressed as
\begin{eqnarray}
G_{i m p}^{(\gamma)}\left( i\omega_n\right)=G_{\gamma}^{(+)}\left( i\omega_n\right)+G_{\gamma}^{(-)}\left( i\omega_n\right),
\end{eqnarray}
where
\begin{eqnarray}
G_{\gamma}^{(+)}\left( i\omega_n\right)=\frac{\left\langle\phi_{0}\left|\gamma \gamma^{\dagger}\right| \phi_{0}\right\rangle}{ i\omega_n-a_{0}^{(+)}-\frac{b_{1}^{(+) 2}}{ i\omega_n-a_{1}^{(+)}-\frac{b_{2}^{(+) 2}}{ i\omega_n-a_{2}^{(+)}-\ldots}}},
\end{eqnarray}

\begin{eqnarray}
G_{\gamma}^{(-)}\left( i\omega_n\right)=\frac{\left\langle\phi_{0}\left|\gamma^{\dagger} \gamma\right| \phi_{0}\right\rangle}{ i\omega_n+a_{0}^{(-)}-\frac{b_{1}^{(-) 2}}{ i\omega_n+a_{1}^{(-)}-\frac{b_{2}^{(-) 2}}{ i\omega_n+a_{2}^{(-)}-\ldots}}}.
\end{eqnarray}

The DMFT simulate lattice model with impurity model
through the self-consistent equation of impurity Weiss
function and the noninteracting Green's function of
lattice model. Considering the semicircular DOS of
the Bethe lattice, the on-site component of the Green's
function of each orbital
$[G^{(\gamma)}_{ii\sigma}(i\omega_n)=\sum_k  G^{(\gamma)}_{\sigma}\left(i\omega_n,k\right)]$
satisfies a simple self-consistent relation\cite{Georges-1996},
\begin{eqnarray}
\left\{g_{0}^{(\gamma)}\left(i\omega_n\right)\right\}^{-1}=i\omega_n+\mu-t_{\gamma}^{2} G_{i m p}^{(\gamma)}\left( i\omega_n\right)
\label{Eq:self-consistent}
\end{eqnarray}
where $g_{0}$ is the noninteracting Green's function
of lattice model. Adjusting the parameters
$V_{m\sigma}^{\gamma}$ and $\epsilon_{m\sigma}^{\gamma}$
to make the impurity Weiss function ${\cal G}_{0}$ equal
with lattice model $g_{0}$, the process of DMFT is complected.

We calculate the orbital-resolved spectral density
of the effective orbital
$\gamma$ by
\begin{equation}
A_{\gamma}(\omega)=-\frac{1}{\pi}\rm{Im}\it{G^{\gamma}_{ii}(\omega+i\eta)},
\end{equation}
where $\eta$ is an energy broadening factor.
The orbital-dependent quasiparticle weight
is determined by the self-energy\cite{Liebsch-2007},
\begin{equation}
Z_{\gamma}=(1-\frac{\partial}{\partial\omega}\rm{Re}
\it{\Sigma_{\gamma}(\omega+i\eta)|_{\omega=0}})^{-1}.
\end{equation}

\section{Results and Discussions}

\begin{figure}[htbp]
\includegraphics[scale=0.50]{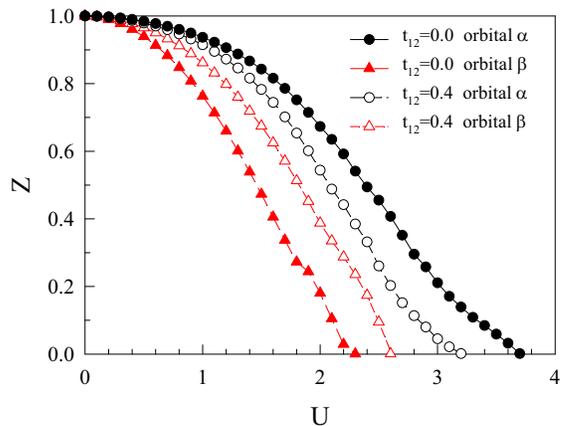}
\caption{(Color online)
The interaction dependencies of the quasiparticle weight
Z of the effective orbital $\alpha$ (black circles) and $\beta$ (red triangles) with interorbital hopping $t_{12}=0.4$ (hollow symbols) and without interorbital hopping $t_{12}=0$ (solid symbols) when $J_H=0.25U$ and $t_2/t_1=0.4$.
The energy is in unit $t_1$.
}
\label{fig:ZU-t12}
\end{figure}

We study the effects of interaction and interorbital
hopping on phase transition in the extended two-orbital Hubbard model. The chemical potential $\mu$ is kept as
$\mu=U/2+U^{\prime}-J_H/2$ to satisfy the particle-hole
symmetry.
We compare in Fig.~\ref{fig:ZU-t12} the quasiparticle
weights for different interaction $U$ and interorbital
hopping $t_{12}$.
While $t_{12}=0$, the quasiparticle weight $Z$ denotes
that the critical interaction of metal-insulator transition (MIT) $U_{c\alpha}=3.7$ for the wide effective orbital $\alpha$, and the critical interaction for
the narrow effective orbital $\beta$ is $U_{c\beta}=2.3$, indicating that the OSMP occurs when
interaction $2.3\leq U< 3.7$.
The OSMP region of effective model narrows to $2.6\leq U< 3.2$
when $t_{12}=0.4$ as shown in
Fig.~\ref{fig:ZU-t12}, where the critical point of MIT
for orbital $\alpha$ shifts to the weak
interaction region and the narrow orbital critical
interaction becomes strong. The positive
interorbital hopping ($t_{12}>0$) suppresses the region of OSMP by
decreasing the difference of two effective orbital hopping integrals, i.e. $t_{\beta}/t_{\alpha}=0.67$ according to Eq.~(\ref{Eq:CT}), which is opposite to previous results for the negative interorbital hopping that enhances the OSMP\cite{Song-2009}.

%++++++++++++++++++++++++%
\begin{figure}[htbp]
\includegraphics[scale=0.50]{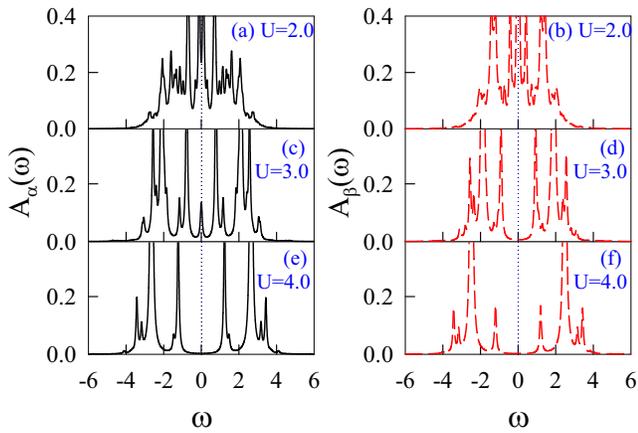}
\caption{(Color online)
Evolution of the orbital-resolved spectral density $A(\omega)$ with the increasing Coulomb
interaction $U$ when $t_{12}=0.4$, $t_2/t_1=0.4$ and $J_H=0.25U$.
Left panels show the results of effective orbital $\alpha$ and right panels are for
effective orbital $\beta$. The energy boarding factor in our calculation takes $\eta=0.05$.}
\label{fig:DOS-t4}
\end{figure}

Fig.~\ref{fig:DOS-t4} shows the details of spectral density evolution
with increasing interaction $U$ for
$t_{12}=0.4$ when $t_2/t_1=0.4$ and $J_H=0.25U$.
At the condition of $U=2.0$, there exist
resonance peaks around the Fermi level for both effective
orbital $\alpha$ [Fig.~\ref{fig:DOS-t4} (a)] and
orbital $\beta$ [Fig.~\ref{fig:DOS-t4} (b)]. The
finite spectral weights indicate that both
effective orbitals are metallic, so that the system is in metal phase.
When interaction $U=3.0$, the spectral weight
at the Fermi level of orbital $\alpha$ is finite
[Fig.~\ref{fig:DOS-t4} (c)], but a Mott gap opens around
the Fermi level in orbital $\beta$ [Fig.~\ref{fig:DOS-t4} (d)].
This is the typical characteristic of OSMP. Increasing
interaction to $U=4.0$, Mott gaps can be found in the
spectral density of both bands in Fig.~\ref{fig:DOS-t4}
(e) and (f), and then the system transforms into insulating
phase.

\begin{figure}[htbp]
\includegraphics[scale=0.43]{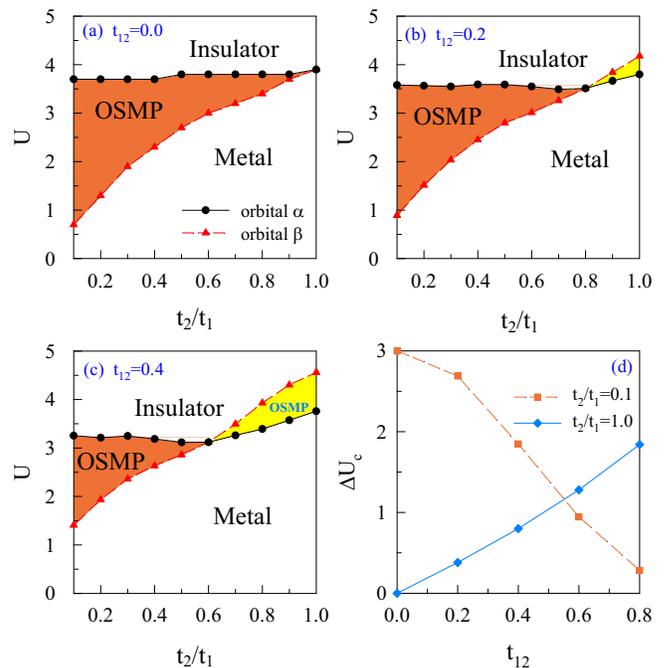}
\caption{(Color online)
Phase diagrams of the effective two-orbital Hubbard model with different interorbital hopping:
(a) $t_{12}=0.0$, (b) $t_{12}=0.2$ and (c) $t_{12}=0.4$, when $J_H=0.25U$. The black circles (red triangles) denote the critical points of MIT for effective orbital $\alpha$ ($\beta$).
(d) The $t_{12}$ dependence
of the difference between the critical interactions of the two effective orbitals ($\Delta U_{c}=\left|U_{c \alpha}-U_{c \beta}\right|$) for $t_2/t_1=0.1$ (orange square symbol) and $t_2/t_1=1.0$ (blue diamond symbol).}
\label{fig:PD-compare}
\end{figure}

We construct phase diagrams in the plane of
interaction $U$ and hopping integral $t_2/t_1$
with different interorbital hopping
in Fig.~\ref{fig:PD-compare}. In a two-orbital
system without interorbital hopping, as shown in
Fig.~\ref{fig:PD-compare} (a), the OSMP can exist for any
hopping integral ratio except $t_2/t_1=1$ because of the effect
of Hund's rule coupling, and the OSMP region narrows
with increasing $t_2/t_1$, which is consistent
with the pervious research on OSMP\cite{Koga-2004,Costi-2007,Medici-2011}.
When we introduce the interorbital hopping, the hopping ratio of two effective orbitals will increase, as a surelt the normal OSMP region (orange region) shrinks as shown in Fig.~\ref{fig:PD-compare} (b) and (c) with $t_{12}=0.2$ and $t_{12}=0.4$ respectively.
It is worth noting that a new OSMP (yellow region) appears and its region expands with the increasing interorbital hopping.
In the new OSMP, the effective
orbital $\alpha$ translates into insulator while orbital
$\beta$ keeps in metal phase.%

In order to exhibit the change of OSMP region directly, we show the $t_{12}$ dependence of $\Delta U_{c}$ in Fig.~\ref{fig:PD-compare} (d). When $t_2/t_1=0.1$, $\Delta U_{c}$ (orange square symbol) decreases as $t_{12}$ increases, which denotes the normal OSMP region gradually shrinks under the effect of $t_{12}$.
Conversely, the rising blue curve denotes the new OSMP region expends with increasing $t_{12}$ when $t_2/t_1=1.0$.
The effective hopping integral ratio $t_{\beta}/t_{\alpha}$
increases with increasing $t_2/t_1$ under the effect of interorbital
hopping $t_{12}$. When $t_{12}=0.2$ and $t_2/t_1=0.8$, the effective
orbital hopping $t_\alpha=t_\beta=0.70$ according to
Eq.~(\ref{Eq:CT}). If $t_2/t_1>0.8$, $t_{\beta}$ is greater
than $t_{\alpha}$, thus the new OSMP region appears
in Fig.~\ref{fig:PD-compare} (b). Correspondingly, $t_\alpha=t_\beta=0.80$ when
$t_{12}=0.4$ and $t_2/t_1=0.6$. It causes the movement of intersection and the change of OSMP region in Fig.~\ref{fig:PD-compare} (c).

\begin{figure}[htbp]
\includegraphics[scale=0.50]{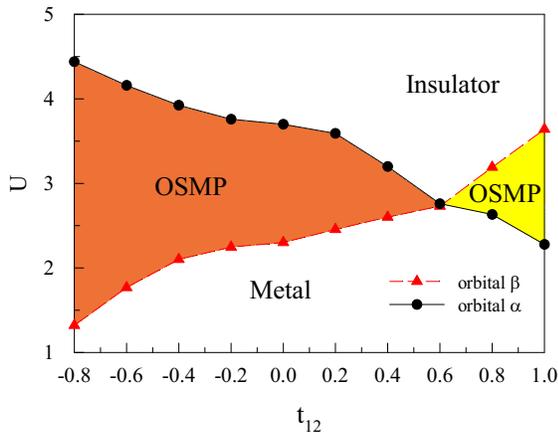}
\caption{(Color online)
Phase diagram under the effect of interaction $U$ and interorbital hopping $t_{12}$
when $J_H=0.25U$ and $t_2/t_1=0.4$. Two OSMP regions are found in the phase diagram.}
\label{fig:PD-t12}
\end{figure}

Electronic structures of solid materials are complicated and various. The phase difference between the plane wave of electron in different orbitals may make the interorbital hopping integral be negative\cite{Ernzerhof-1996,Madsen-2002,Held-2010}. Thus we extend the interorbital hopping to $-0.8\leq t_{12}\leq1.0$,
and construct the phase diagram under the effect of
interorbital hopping in Fig.~\ref{fig:PD-t12}.
Two interleaved OSMP regions
separate the metal phase and insulator phase. At the
region of $t_{12}>0$, normal OSMP region decreases accompanying
with increasing $t_{12}$ until $t_{12}=0.6$.
It transforms into a new OSMP while $t_{12}>0.6$.
The physical mechanism is consistent with the
Fig.~\ref{fig:PD-compare} (b) and (c).
Fig.~\ref{fig:PD-t12} also shows the OSMT while $t_{12}<0$,
the normal OSMP region expands with the increasing $\mid t_{12} \mid$
indicating negative interorbital hopping integral $t_{12}$ is beneficial
to normal OSMP. The interorbital hopping will
increase the difference of two effective orbital hopping integrals
if $t_{12}<0$ according to Eq.~(\ref{Eq:CT}), thus the
multi-orbital character will be enhanced under the
effect of correlation. As a result, the normal OSMP region enlarges
in the phase diagram.

%^^^^^^^^^^^^^^^^^^^^^^^^^^^^^^^^^^^^^^^^^%
\section{Application}

In this section, we apply the extended DMFT in recently discovered
superconductor Ba$_2$CuO$_{4-\delta}$\cite{JinCQ-PNAS-2019}, which can be
described with two-orbital Hubbard model.
Based on the DFT-calculated band structure of the compressed
half-filled Ba$_2$CuO$_{3.5}$ compound, the electronic states near the Fermi level consist primarily of the Cu $d_{x^2-y^2}$ and $d_{3z^2-r^2}$ orbitals.
The model parameters of the tight-binding Hamiltonian $H_t$
in Eq.~(\ref{Eq:tightbonding}) take the following values:
$t_{1}$=0.504~eV, $t_{2}$=0.196~eV, and $t_{12}=-0.302$~eV. It is obvious that this compond belongs to the two-orbital system with a negative interorbital hopping.
As discussed in the above section, the negative interorbital hopping is favorable to the existence of the OSMP, thus we predict that Ba$_2$CuO$_{4-\delta}$ is an OSMP compound.
A crystal-field splitting $\epsilon_d=\mu_1-\mu_2$ is introduced with $\mu_1=-0.222$~eV, and $\mu_2=0.661$~eV
\cite{Scalapino-2019}, so that the particle-hole symmetry is broken in Ba$_2$CuO$_{3.5}$ band structure.
A constant in off-diagonal part of the Green's function matrix induced by the crystal-field splitting should be also considered, and the details of the modified DMFT procedure can be found in reference \cite{Song-2009}.

\begin{figure}[htbp]
\includegraphics[scale=0.50]{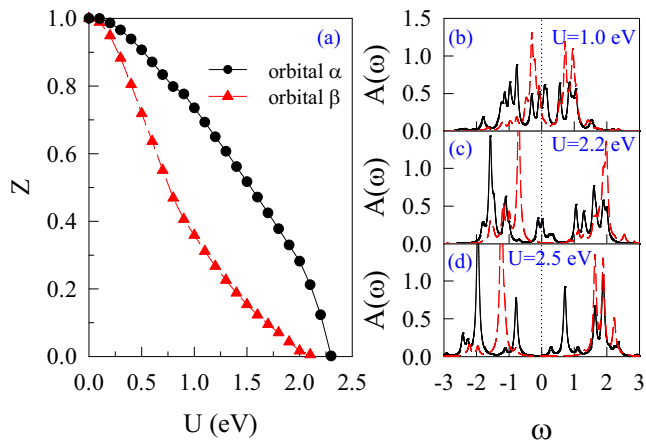}
\caption{(Color online)
  (a) Quasiparticle weight $Z$ as a function of interaction $U$ when $J_H=0.25U$ for Ba$_2$CuO$_{3.5}$. The orbital-resolved spectral
  density $A(\omega)$ with different intraorbital interaction: (b) $U=1.0$~eV, (c) $U=2.2$~eV, and (d) $U=2.5$~eV for the two effective orbitals.
  An OSMP occurs in a narrow interaction region with 2.1 eV $\leq U<$ 2.3 eV.
}
\label{fig:ZA-J4}
\end{figure}

The orbital-dependent quasiparticle weight $Z_{\gamma}$ as
a function of the interaction $U$ when $J_H=0.25U$
is shown in Fig.~\ref{fig:ZA-J4} (a).
According to quasiparticle weight $Z$, the two-orbital system is metallic
when $U<2.1$~eV, and it transforms into insulator while $U\geq 2.3$~eV.
A narrow OSMP region exists within $2.1~$eV$\leq U<2.3~$eV, in which
the wide $\alpha$ band is metallic and
the narrow $\beta$ band behaves insulating.
Fig.~\ref{fig:ZA-J4} (b) (c) and (d) reveal the effects of interaction $U$ on the
orbital-resolved spectrum $A(\omega)$
for both the effective $\alpha$ and $\beta$ bands in Ba$_2$CuO$_{3.5}$
with $J_H=0.25U$. When the interaction is weak, such as
$U=1.0$~eV, the spectral weights at the Fermi level are finite for both bands, as
shown in Fig.~\ref{fig:ZA-J4} (b), indicating
a metallic phase for the two-orbital system.
With increasing interaction $U$ to 2.5~eV, Mott gaps can be found in the
DOS of both bands in Fig.~\ref{fig:ZA-J4} (d), thus the system is insulator at this condition.
When $U=2.2$ eV as
shown in Fig.~\ref{fig:ZA-J4} (c), the wide $\alpha$ band is
still metallic with the resonance peaks in its DOS at the Fermi level,
but a Mott gap around the Fermi level exists in the narrow $\beta$ band
,indicating the system is in the
OSMP\cite{JSun-2015,Costi-2007,Medici-2011}.

%++++++++++++++++++++++++%
\begin{figure}[htbp]
\includegraphics[scale=0.50]{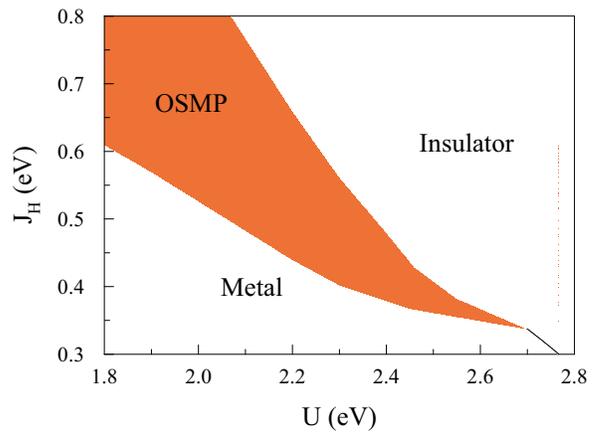}
\caption{(Color online) The phase diagram of the effective two-orbital
Hubbard model with interaction $U$ and Hund's rule coupling $J_H$
for Ba$_2$CuO$_{3.5}$.
%Both the critical values $U_{c \beta}$ and $U_{c\alpha}$ decrease with the increasing $J_H/U$.
The region of OSMP becomes narrower with the decreasing
$J_H$ and increasing $U$, and OSMT vanishes around $J_H=0.34$~eV and $U=2.7$ eV.}
\label{fig:JU-HF}
\end{figure}

The phase diagram of Ba$_2$CuO$_{3.5}$
in the plane of interaction $U$ and Hund's rule coupling $J_H$
is shown in Fig.~\ref{fig:JU-HF}.
Between the strongly correlated Mott insulating phase and weak correlated metallic phase,
the OSMP region shrinks accompanying with decreasing $J_H$ and increasing $U$, and vanishes while $J_H=0.34$~eV and $U=2.7$~eV.
Hund's rule coupling $J_H$ is beneficial
to the occurrence of the OSMP according to precious results\cite{Liebsch-2005,deMedici-2011},
which can explain the evolutionary trend of OSMP in the large $J_H$ region.
When $J_H<0.34$~eV, the crystal-field splitting is superior to
Coulomb correlation and Hund's rule coupling, so that OSMP vanishes from phase diagram\cite{Liebsch-2005,deMedici-2011,Georges-2013}.
Ba$_2$CuO$_{3.5}$ belongs to transition-metal oxides,
in which the electron-electron correlation caused by $d$-electron
is considerable, so it should be an OSMP compound predicted by our calculation.

%*****************************%

%^^^^^^^^^^^^^^^^^^^^^^^^^^^^^^^^^^^^^^^^^%
\section{conclusions}

To conclude, we study the effects of interaction and interorbital
hopping on the OSMT in two-orbital Hubbard model by using
DMFT with Lanczos solver, and we find that the interorbital hopping influences phase transition by changing
the hopping integral ratio of effective orbitals: (1) if the interorbital hopping $t_{12}>0$, the wide band becomes narrow but the narrow band is getting broaden with increasing $t_{12}$,
and the OSMP is suppressed until $t_{\alpha}=t_{\beta}$.
Conversely, a new OSMP appears when we increase $t_{12}$ continuously, where the
effective orbital $\beta$ becomes metallic and orbtial $\alpha$ behaves
as insulator; (2) if $t_{12}<0$, interorbital hopping enhances the
OSMP by increasing the difference of the two effective orbitals.
We apply the extended DMFT method to construct the phase diagram
of the recently discovered
two-orbital superconductor Ba$_2$CuO$_{3.5}$, and we demonstrate
that the half-filled Ba$_2$CuO$_{3.5}$ should be an OSMP
compound.

\section*{Acknowledgments}
We thanks Liang-Jian Zou for helpful discussions.
This work is supported by the National Natural Science Foundation of China (NSFC)
under the Grant Nos. 11474023, 11774350 and 11174036.


\end{document}